\documentclass[amssymb,epsfig,color,11pt]{article}
\usepackage{epsfig}
\usepackage{amsfonts}
\usepackage{amssymb}
\usepackage{indentfirst}
\usepackage{color}
\usepackage{mathrsfs}
\def\bc{\begin{center}}

\def\ec{\end{center}}
\def\be{\begin{eqnarray}}
\def\ee{\end{eqnarray}}

\definecolor{dyellow}{rgb}{1.,0.8,.0}
\definecolor{myblue}{rgb}{.1,.1,.7}
\definecolor{dcyan}{rgb}{.0,.6,.6}
\definecolor{dmagenta}{rgb}{0.6,0.0,0.6}
\definecolor{brown}{rgb}{0.6,0.2,0.}
\definecolor{darkblue}{rgb}{.0,.0,0.5}
\definecolor{darkred}{rgb}{0.75,0.0,0.0}
\definecolor{orange}{rgb}{1.,.6,.0}
\definecolor{dorange}{rgb}{0.8,.4,.0}
\definecolor{darkgreen}{rgb}{0.0,0.6,0.0}
\definecolor{purple}{rgb}{.4,.0,.4}

\begin{document}
\baselineskip=16pt
\newcommand{\omits}[1]{}

\hfill{\bf USTC-ICTS-07-05} \vspace{1.0cm}

\begin{center}
{\Large \bf Variation of the Fine-Structure Constant from the de Sitter Invariant Special Relativity}\\
\vspace{1cm}
Shao-Xia Chen, Neng-Chao Xiao, Mu-Lin Yan\footnote{mlyan@ustc.edu.cn}\\
{\em Interdisciplinary Center for Theoretical Study,} \\
{\em University of
Science and Technology of China,}\\
{\em Hefei, Anhui 230026, China}\\
\end{center}

\begin{abstract}
There are obvious discrepancies among various experimental
constraints on the variation of the fine-structure constant,
$\alpha$. We attempt to discuss the issue in the framework of de
Sitter invariant Special Relativity (${\cal SR}_{c,R}$) and to
present a possible solution to the disagreement. In addition, on
the basis of the observational data and the discussions presented
in this Letter, we derive a rough theoretical estimate of the
radius of the Universe.
\end{abstract}

\vspace{1.0cm}


\vspace{1.0cm}

The issue of a possible variation of fundamental physical
constants has been put on the agenda of contemporary physics.
Recently, Webb group reported \cite{webb06} a varying
fine-structure constant $\alpha$ through analyzing the multiple
heavy element transitions in the absorption spectra of
quasi-stellar objects (QSOs) on the Southern Hemisphere, which
agrees with and offers vigorous support to their previous research
results
\cite{webb06,Murphy03b,Murphy03a,murphy01b,murphy01a,webb01,webb99}
on the Northern Hemisphere: Compared with laboratory data on
atomic multiplet structures, the observations of the absorption
lines of QSOs reveal that the fine-structure constant $\alpha$ has
an obvious change in the past several billion years. This
confirmation stimulates the further study on the variation of
fine-structure constant both from the theoretical interpretation
\cite{theo1,theo2} to establish the motivation and to supply more
exact model analysis, and the experimental measurements
\cite{experiment} to yield more precise data and more experimental
methods.

The possibility of variability of fundamental constants was put
forward by Dirac \cite{dirac,peng} in 1937, afterwards, a lot of
theoretical illustration and experimental constraints
\cite{review} on the variation of fundamental constants are
presented. As we know, the constancy of the fundamental constants
plays a significant role in astronomy and cosmology where the
redshift measures the look-back time. If ignoring the possibility
that the constants are varying we will have a deviated view of our
universe. However, if such variations are established, corrections
should be applied to the related issues. It is thus necessary to
investigate that possibility, especially as the measurements
become more precise and/or when the measurements are made on the
larger scale.

Besides, a general feature of extra-dimensional theories, such as
Kaluza-Klein and string theories, is that the ``true" fundamental
constants of nature are defined in the full higher dimensional
theory, so that the effective 4-dimensional ``fundamental
constants'' depend, among other things, on the structure and sizes
of the extra-dimensions. The time and/or space evolution  of these
sizes would have a significant result that the effective
4-dimensional ``fundamental constants'' will also depend on the
spacetime. What's more, the achievement of experimental
constraints on the variation of fundamental constants is dependent
on the high energy physics models, to some extent. For two
aforementioned reasons, the observation of the variability of
fundamental constants is one of the few ways to test directly the
existence of extra-dimensions and to test high energy physics
models.

Among all the possible fundamental constants of nature we will
focus on the fine-structure constant, $\alpha$, which can be
derived from other constants as follows \cite{alpha0}:
\begin{equation}\label{define}
  \alpha=\frac{e^2}{4\pi\hbar c}~,
\end{equation}
where $c$ is the speed of light, $\hbar\equiv h/2\pi$ is the
reduced Planck constant, $e$ is the electron charge magnitude. The
value of $\alpha$ measured today on earth is
$\alpha_0\approx1/137.035$ \cite{alpha0}.

As the experimental constraints on other fundamental constants,
there is also inconsistence among different groups measuring the
change of $\alpha$, especially between the nonzero observational
results
\cite{webb06,Murphy03b,Murphy03a,murphy01b,murphy01a,webb01,webb99}
from the absorption lines of QSOs and the null bounds
\cite{Oklo1,Oklo2,Oklo3,Oklo4,Oklo5,Oklo6} from the capture cross
section of thermal neutron by $_{62}^{149}$Sm in the natural
fission reactor that operated about $2\times10^9$yr ago during
$(2.3\pm0.7)\times10^5$yr in the Oklo uranium mine in Gabon.
Dirac's large numbers hypothesis \cite{dirac} conjectured that the
fundamental constants are functions of the epoch, so it is clear
the hypothesis is not able to disentangle the inconsistence in the
different observations on its own.

The recent astronomical observations on supernovae
\cite{constant-super-1,constant-super-2} and CMBR
\cite{costant-cmbr} manifest that about $73\%$ of the whole energy
in the Universe is dark energy and possibly contributed by a tiny
positive cosmological constant $(\Lambda)$. The observations
strongly indicate that the Universe is described by an
asymptotically de Sitter (dS) spacetime, which stimulates the
interests to reconsider the de Sitter invariant special relativity
with two universal constants $c$ and $R$, which is shortly denoted
as ${\cal SR}_{c,R}$ \cite{dS-SR1,dS-SR2,dS-SR3,dS-SR4,yan}.

One basis to establish ${\cal SR}_{c,R}$ is the principle of
special relativity: There exist a set of inertial reference
frames, in which the free particles and light signals move with
uniform velocity along straight lines, {\it i.e.}, the inertial
motion law  holds true in the frames. The other is the postulate
of invariant constants: There exist two invariant universal
constants, {\it i.e.}, speed $c$ and length $R$. The key point to
set up ${\cal SR}_{c,R}$ is that in dS spacetime there is an
important kind of coordinate system, called the Beltrami
coordinate system, which is analogous to the Minkowski one in a
flat spacetime. In the Beltrami coordinate system, test particles
and light signals move along the time-like and null geodesics,
respectively, with constant coordinate velocity. Therefore,
inertial observers and classical observable quantities for these
particles and signals in the Beltrami coordinate system can be
well defined.

Obviously, the fine-structure constant has not undergone huge
variations on Solar system scales and on geological time scales,
so one is looking for measurable effects on the larger scale, such
as astrophysical and astronomical scale, even cosmological one. In
the meanwhile, it is also necessary for the experimental tests to
be carried on on the large scale for the purpose of distinguishing
${\cal SR}_{c,R}$ from Einstein's Special Relativity. These
considerations motivate us to investigate the possibility of
making use of ${\cal SR}_{c,R}$ to illustrate the inconsistence
between the observations of variation of $\alpha$ in QSOs and in
the Oklo natural reactor.

In this Letter, we address the issue of variation of the
fine-structure constant in the framework of ${\cal SR}_{c,R}$ in
order to examine the possibility of solving the inconsistence
between the observational constraints from the absorption spectra
of QSOs and from the Oklo phenomenon. In addition, through
comparing the theoretical derivation and the observational data,
we crudely provide the value of the radius of the Universe, $R$.

The dS spacetime can be realized as a four dimensional
hypersurface embedded in a five dimensional flat space
\begin{equation}
ds^2=(d\xi^0)^2-(d\xi^1)^2- (d\xi^2)^2-(d\xi^3)^2-(d\xi^4)^2~,
\end{equation}
such that
\begin{equation} \label{hypersurface}
(\xi^0)^2-(\xi^1)^2-(\xi^2)^2-(\xi^3)^2 -(\xi^4)^2=-R^2~.
\end{equation}
The metric of the 4-dimensional spacetime in Beltrami coordinates
has the form
\begin{equation}\label{metric}
\begin{array}{rcl}
ds^2&=&g_{\mu\nu}dx^\mu dx^\nu~,\\ [0.3cm]  g_{\mu\nu}&=&
\displaystyle\frac{\eta_{\mu\nu}} {\sigma(x,x)}+\frac{\eta_{\mu
\alpha}\eta_{\nu \beta}x^{\alpha}x^{\beta}}
{\sigma(x,x)^2R^2}~,\\[0.5cm]
\sigma(x,y)&=&1-\displaystyle\frac{\eta_{\mu\nu}x^{\mu}y^{\nu}}{R^2}>0~,
\\ [0.5cm] \eta_{\mu\nu}&=&{\rm diag}(1,~-1,~-1,~-1)~,
~~~~\alpha,~\beta,~\mu,~\nu=0,~1,~2,~3.
\end{array}
\end{equation}
It is invariant under the fractional linear transformation
\begin{equation}\label{trans-gen}
\begin{array}{rcl}
 x^\mu\rightarrow\bar{x}^\mu&=&
 \displaystyle\frac{\sqrt{\sigma(a)}}{\sigma(a,x)}D^\mu_\nu(x^\nu-a^\nu)~,
 \\ [0.3cm] D^\mu_\nu&=&L^\mu_\nu+\displaystyle\frac{1}{R^2}
 \displaystyle\frac{L^\mu_\lambda a^\lambda\eta_{\nu\rho}a^\rho}
 {\sigma(a)+\sqrt{\sigma(a)}}~,\\ [0.5cm]
L^\mu_\nu&\in&SO(1,3)~,
\\ [0.2cm] \sigma(a)&\equiv&\sigma(a,a)=1-\displaystyle\frac{\eta_{\mu\nu}a^\mu a^\nu}{R^2}~.
\end{array}
\end{equation}
Here, $(a^0,~a^1,~a^2,~a^3)$ is the spacetime coordinate of the
origin of the resulted inertial Beltrami frame in the original
inertial Beltrami frame, and we note we use the units in which the
speed of light in vacuum in Einstein's Special Relativity is
$c=1$.

Considering the physical issue we investigate, we can suppose that
there aren't relative motion and space rotations between two
inertial Beltrami frames, and we can also reduce our discussions
into two dimensions, {\it i.e.}, we only translate the spacetime
origin in $x^0-x^1$ plane, $a^2=a^3=0$. Under the assumptions, the
coordinate transformation between two inertial frames in Betrami
de Sitter spacetime is
\begin{equation}
\label{x-trans}
 \begin{array}{rcl}
  x^0\to\bar{x}^0&=&\displaystyle\frac{\sqrt{\sigma(a)}}{\sigma(a,x)}
  \left[x^0-a^0+\displaystyle\frac{a^0}{R^2}
  \displaystyle\frac{a^0x^0-a^1x^1+(a^1)^2-(a^0)^2}{\sqrt{\sigma(a)}+\sigma(a)}\right]~,
  \\ [0.5cm]
x^1\to\bar{x}^1&=&\displaystyle\frac{\sqrt{\sigma(a)}}{\sigma(a,x)}
  \left[x^1-a^1+\displaystyle\frac{a^1}{R^2}
  \displaystyle\frac{a^0x^0-a^1x^1+(a^1)^2-(a^0)^2}{\sqrt{\sigma(a)}+\sigma(a)}\right]~,
  \\ [0.5cm]
x^2\to\bar{x}^2&=&\displaystyle\frac{\sqrt{\sigma(a)}}{\sigma(a,x)}x^2~,
\\[0.5cm]
x^3\to\bar{x}^3&=&\displaystyle\frac{\sqrt{\sigma(a)}}{\sigma(a,x)}x^3~.
 \end{array}
\end{equation}

The energy and momentum of a photon (or light signal) is defined
as \cite{huang}
\begin{equation}\label{private}
  E \equiv\epsilon~\displaystyle\frac{dx^0}{d\lambda}~,
  ~~~~~~~~~~~~~~~~~p^i \equiv\epsilon~\displaystyle\frac{dx^i}{d\lambda}~,
\end{equation}
where $\lambda$ is an affine parameter and $\epsilon$ is a
constant. The velocity of light is the ratio of the energy to the
momentum of a photon
\begin{equation}\label{speed-light}
  \tilde{c}\equiv\displaystyle\frac{E}{p}~,
\end{equation}
and the corresponding physical fine-structure constant
becomes
\begin{equation}\label{define1}
  \alpha=\frac{e^2}{4\pi\hbar \widetilde{c}}~.
\end{equation}
From the coordinate transformation Eq.(\ref{x-trans}), it is easy
to derive the coordinate velocity transformation between two
inertial Beltrami frames. For the velocity of light in the $x^1$
direction through the origin of the original frame, in the
transformed one, it reads
\begin{equation}\label{u-trans}
     \tilde{c}=\frac
     {1-\frac{(a^0)^2}{R^2}}{\sqrt{\sigma(a)}-\frac{a^0a^1}{R^2}}~.
\end{equation}
Examining the QSO frame and the earth frame, we have $a^0=a^1$,
therefore,
\begin{equation}\label{QSO-trans}
     \tilde{c}_{\rm QSO}=1~.
\end{equation}
It means that the speed of light emitted from QSO to the earth is
the same as that observed today on the earth.

In the meanwhile, the QSO observations show that fine-structure
constant has a nonzero change
$\Delta\alpha/\alpha_0\equiv(\alpha_{\rm
past}-\alpha_0)/\alpha_0\sim-10^{-5}$, which can only come from
the variation of the part $e^2/\hbar$.

The large numbers hypothesis is raised by Dirac \cite{dirac},
which argued the fact that some large dimensionless numbers have
the same order leads one to believe some fundamental constants
vary with the epoch. Based on this hypothesis, we assume
$e^2/\hbar$ is only the function of time $t$, so the variations of
$\alpha$ in the QSO observations are the results due to the
variations of $e^2/\hbar$, that is
\begin{equation}
\label{QSO-e}
\begin{array}{rcl}
 \left.\displaystyle\frac{\Delta\alpha}{\alpha}\right|_{\rm QSO}\equiv
   \displaystyle\frac{\alpha_{\rm QSO}-\alpha_0}{\alpha_0}
&=&\displaystyle\frac{\displaystyle\frac{1}{4\pi}\left(\displaystyle\frac{e^2}{\hbar}\right)_{\rm
QSO}\times\left(\displaystyle\frac{1}{\tilde{c}}\right)_{\rm
QSO}-\displaystyle\frac{1}{4\pi}\left(\displaystyle\frac{e^2}{\hbar}\right)_0
\times\left(\displaystyle\frac{1}{\tilde{c}}\right)_0}{\displaystyle\frac{1}{4\pi}\left(\displaystyle\frac{e^2}{\hbar}\right)_0
\times\left(\displaystyle\frac{1}{\tilde{c}}\right)_0}
 \\ [0.9cm] &=&\displaystyle\frac{\left(\displaystyle\frac{e^2}{\hbar}\right)_{\rm
QSO}-\left(\displaystyle\frac{e^2}{\hbar}\right)_0}
{\left(\displaystyle\frac{e^2}{\hbar}\right)_0}\sim-10^{-5}~,
\end{array}
\end{equation}
thus
\begin{equation}\label{QSO-1}
\left(\displaystyle\frac{e^2}{\hbar}\right)_{\rm
QSO}\sim(1-10^{-5})\times\left(\displaystyle\frac{e^2}{\hbar}\right)_0~.
\end{equation}
In the Letter, the quantities with subscript $0$ stand for those
measured on the earth now.

In the Oklo case, the inertial Beltrami coordinate transformation
is between the present earth and the frame whose origin is the
spacetime point when and where the Oklo phenomenon took place, so
$a^1=0$, and Eq. (\ref{u-trans}) has the form
\begin{equation}\label{Oklo-trans}
\tilde{c}_{\rm Oklo}=\sqrt{1-\frac{(a^0)^2}{R^2}}~.
\end{equation}
Since the Oklo phenomenon occurred before about $2\times10^9$
years, we can cursorily take
\begin{equation}\label{equal}
  \left(\displaystyle\frac{e^2}{\hbar}\right)_{\rm
QSO}=\left(\displaystyle\frac{e^2}{\hbar}\right)_{\rm Oklo}~.
\end{equation}
The Oklo constraints on the variation of fine-structure constant
have null results \cite{Oklo1,Oklo2,Oklo3,Oklo4,Oklo5,Oklo6},
therefore
\begin{equation}\label{oklo-var}
 \left.\displaystyle\frac{\Delta\alpha}{\alpha}\right|_{\rm Oklo}\equiv
   \displaystyle\frac{\alpha_{\rm Oklo}-\alpha_0}{\alpha_0}
=\displaystyle\frac{\displaystyle\frac{1}{4\pi}\left(\displaystyle\frac{e^2}{\hbar}\right)_{\rm
Oklo}\times\left(\displaystyle\frac{1}{\tilde{c}}\right)_{\rm
Oklo}-\displaystyle\frac{1}{4\pi}\left(\displaystyle\frac{e^2}{\hbar}\right)_0
\times\left(\displaystyle\frac{1}{\tilde{c}}\right)_0}{\displaystyle\frac{1}{4\pi}\left(\displaystyle\frac{e^2}{\hbar}\right)_0
\times\left(\displaystyle\frac{1}{\tilde{c}}\right)_0}=0~.
\end{equation}
Combining Eq.(\ref{QSO-1}), Eq.(\ref{equal}) and
Eq.(\ref{oklo-var}), we obtain
\begin{equation}\label{result-1}
\tilde{c}_{\rm
Oklo}=\displaystyle\frac{~~~\left(\displaystyle\frac{e^2}{\hbar}\right)_{\rm
Oklo}}{\left(\displaystyle\frac{e^2}{\hbar}\right)_0}\sim1-10^{-5}~.
\end{equation}
Comparing theoretical calculation Eq.(\ref{Oklo-trans}) with the
result from experimental data Eq.(\ref{result-1}), we can have the
conclusion preliminarily that ${\cal SR}_{c,R}$ can indeed settle
the inconsistence between the Oklo and the QSO observational
results.

Go a step further, we can estimate the radius of the Universe from
the contrast between Eq.(\ref{Oklo-trans}) and Eq.(\ref{result-1})
roughly
\begin{equation}\label{R}
\sqrt{1-\frac{(a^0)^2}{R^2}}\sim1-10^{-5}\Rightarrow R\sim
2\sqrt{5}\times10^{11}{\rm l.y.}\simeq1.37\times10^{11}{\rm
pc}=1.37\times10^{5}{\rm Mpc}>R_0~.
\end{equation}
Here $R_0\sim 10^4$Mpc is the radius (or horizon) of present
observable Universe. It is worth noticing that Eq.(\ref{R}) shows
${\cal SR}_{c,R}$ is consistent with the available cosmological
observations on the $R_0$.

In conclusion, ${\cal SR}_{c,R}$ is really a candidate of the
solution to the inconsistence between the observational results of
the QSO absorption lines and of the Oklo natural reactor on the
variation of the fine-structure constant, which is very different
from the Einstein's Special Relativity. Furthermore, we obtain a
favorable evidence to ${\cal SR}_{c,R}$ from the contrast of the
theoretical assessment with the observational data, that is, the
radius of the Universe, $R$, is greater than the radius (horizon)
of the observable Universe, $R_0$. It is anticipated that as more
experimental methods are applied and more precise observational
data are obtained, ${\cal SR}_{c,R}$ will be confronted with more
stringent tests, even be proved or disproved. Similarly, as ${\cal
SR}_{c,R}$ develops, we will have deeper insight into its
application to various experiments, including the experiments on
the variation of the fine-structure constant.

\vspace{0.5cm}

\section*{Acknowledgments}

We would like to thank Professor Han-Yin Guo, Chao-Guang Huang,
Zhan Xu, and Xing-Chang Song for their helpful discussions and
suggestions. One of us (MLY) wishes to acknowledge Professor
Huan-Wu Peng for introducing the Dirac large numbers hypothesis to
us. The work is supported in part by National Natural Science
Foundation of China under Grant Numbers 90403021, and by the PhD
Program Funds of the Education Ministry of China under Grant
Number 20020358040.


\end{document}